\documentstyle[preprint,aps,pre]{revtex}
\begin{document}
\tighten
\title{Energy nonequipartition in a sheared granular
mixture}
\author{Vicente Garz\'{o}}
\address{Departamento de F\'{\i}sica, Universidad de Extremadura, E-06071 \\
Badajoz, Spain}
\author{Jos\'e Mar\'{\i}a Montanero}
\address{Departamento de Electr\'onica e Ingenier\'{\i}a Electromec\'anica,\\
Universidad de Extremadura, E-06071 \\
Badajoz, Spain}

\date{\today}
\maketitle

\begin{abstract}

The kinetic granular temperatures of a binary granular mixture in simple 
shear flow are determined from the Boltzmann kinetic theory by using a 
Sonine polynomial expansion. The results show that the temperature ratio is 
clearly different from unity (as may be expected since the system is out of 
equilibrium) and strongly depends on the restitution coefficients as well as 
on the parameters of the mixture. The approximate analytical calculations 
are compared with those obtained from Monte Carlo simulations of the 
Boltzmann equation showing an excellent agreement over the 
range of parameters investigated. Finally, the influence of the temperature 
differences on the rheological properties is also discussed.  
\end{abstract}

\draft
%\vspace{1cm}

\bigskip \narrowtext

\newpage

Although experimental and theoretical studies on granular media have been 
mainly focused on assemblies of identical particles, there appears to be a 
recent growing interest among both theorists and experimentalists in the 
more complicated case of polydisperse systems. In this case, several kinetic 
theory studies in the freely cooling \cite{GD99} and thermostatted steady 
\cite{BT02} states have shown that the kinetic temperatures of each species 
are different. This violation of energy equipartition has been subsequently 
confirmed in experiments of vibrated granular mixtures\cite{FM02}. When the 
system is sheared, a similar result has been recently found by Clelland and 
Hrenya \cite{CH02} from molecular dynamics (MD) simulations of a 
binary-sized mixture of inelastic, smooth hard disks engaged in rapid shear 
flow. Their results were compared with previous kinetic theory 
calculations\cite{JM89} based on the assumption of equipartition of granular 
energy. As Clelland and Hrenya conclude \cite{CH02}, although this 
equipartition-of-energy assumption does not appear to have a negative impact 
on the ability of those earliest theories to predict the stress tensor in 
simple shear flow, a {\em multi-temperature} theory must be more 
appropriate. In the context of kinetic theory, the only primary attempts to 
include temperature differences in dense granular mixtures were put forward 
by Jenkins and Mancini \cite{JM87} and more recently by Huilin {\em et 
al.}\cite{HGM01}. However, both works are phenomenological with no attempt 
to solve the kinetic equation. Instead, they assume that the velocity 
distribution functions are local Maxwellians. This can be reasonable to 
estimate the dense gas collisional transfer contributions to the fluxes, but 
not to compute their kinetic contributions. On the other hand, both theories 
are applicable to a general flow field.

The goal of this brief report is to explicitly get the dependence of 
the temperature ratio $\gamma\equiv T_1/T_2$ on particle properties as 
well as on compositional parameters of a granular binary mixture subjected 
to the simple shear flow. The calculation of $\gamma$ allows one to assess 
the magnitude of the equipartition violation and its dependence on 
the parameters of the system. On the other hand, due to the complexity of 
the general problem, here we will restrict ourselves to the low-density 
regime in which case the velocity distribution functions $f_i$ for the two 
species verify a set of two coupled nonlinear Boltzmann equations. This set 
of equations is first analytically solved by using a first-Sonine polynomial 
approximation. Then, to check the reliability of our theoretical results, 
the direct simulation Monte Carlo method \cite{B94} is also employed to 
numerically solve the Boltzmann equation in the shear flow state. As will be 
shown later, our theory presents a much better agreement with simulations 
than the one given by Jenkins and Mancini \cite{JM87}.

Let us consider a granular binary mixture composed by smooth inelastic disks 
or spheres of masses $m_1$ and $m_2$ and diameters $\sigma_{1}$ and 
$\sigma_{2}$. Collisions between particles are inelastic and characterized 
by three constant (independent) restitution coefficients $\alpha_{11}$, 
$\alpha_{22}$, and $\alpha_{12}=\alpha_{21}$, where $\alpha_{ij}\leq 1$ 
refers to the restitution coefficient for collisions between particles of 
species $i$ and $j$. The mixture is under simple shear flow, namely, a 
macroscopic state with a constant linear velocity profile ${\bf 
U}={\sf a}\cdot {\bf r}$, where $a_{k\ell}=a\delta_{kx}\delta_{\ell y}$, $a$ 
being the constant shear rate. In addition, the partial densities $n_i$ and 
the (global) granular temperature $T$ are uniform. The time evolution of the 
temperature arises from the balance of two competing effects: viscous 
heating and collisional cooling. When both mechanisms cancel each other, the 
system reaches a steady state and the temperature achieves a constant value. 
This steady state is what we want to analyze here. From a microscopic point 
of view, the simple shear flow problem becomes spatially uniform in the 
local Lagrangian frame moving with the flow velocity ${\bf U}$. In this 
frame, $f_i({\bf r},{\bf v})\to f_i({\bf V})$, where ${\bf V}={\bf v}-{\bf 
U}$ is the peculiar velocity. Under these conditions, the set of Boltzmann 
kinetic equations read 
\begin{equation}
\label{1}
-a V_{y}\frac{\partial}{\partial V_{x}}f_i({\bf V})=\sum_jJ_{ij}[{\bf 
V}|f_i,f_j],
\end{equation} 
where the Boltzmann collision operator $J_{ij}\left[{\bf 
V}|f_{i},f_{j}\right]$ describing the scattering of pairs of particles is 
\begin{equation}
\label{2}
J_{ij}\left[{\bf V}_{1}|f_{i},f_{j}\right]=
\sigma _{ij}^{d-1}\int d{\bf V}_{2}\int d\widehat{\bbox {\sigma}}\,\Theta 
(\widehat{\bbox {\sigma}}\cdot {\bf g}_{12})(\widehat{\bbox {\sigma }}\cdot 
{\bf g}_{12}) 
\left[ \alpha_{ij}^{-2}f_{i}({\bf V}_{1}')f_{j}({\bf 
V}_{2}')-f_{i}({\bf V}_1)f_{j}({\bf V}_{2})\right] 
\;. 
\end{equation}
Here, $d$ is the dimensionality of the system, $\sigma_{ij}=\left( 
\sigma_{i}+\sigma_{j}\right)/2$, $\widehat{\bbox {\sigma}}$ is a unit vector 
along their line of centers, $\Theta$ is
the Heaviside step function and ${\bf g}_{12}={\bf V}_{1}-{\bf V}_{2}$. In 
addition, the primes on the velocities denote the initial values $\{{\bf 
V}_{1}^{\prime},{\bf V}_{2}^{\prime}\}$ that lead to $\{{\bf V}_{1},{\bf 
V}_{2}\}$ following a binary collision: ${\bf V}_{1}^{\prime}={\bf 
V}_{1}-\mu_{ji}\left(1+\alpha_{ij}^{-1}\right)(\widehat{\bbox 
{\sigma}}\cdot {\bf g}_{12})\widehat{\bbox {\sigma}}$ and ${\bf 
V}_{2}^{\prime}={\bf V}_{2}+\mu_{ij}\left( 1+\alpha_{ij}^{-1}\right) 
(\widehat{\bbox {\sigma}}\cdot {\bf g}_{12})\widehat{\bbox{\sigma}}$. 
Here, $\mu_{ij}=m_{i}/\left(m_{i}+m_{j}\right)$.

Our study is focused on the evaluation of the partial temperatures $T_i$, 
which measure the mean kinetic energy of each species. In terms of $f_i$, 
they are defined as 
\begin{equation}
\label{3}
\case{d}{2} n_iT_i=\int d{\bf V}\case{1}{2}m_iV^2f_i.
\end{equation} 
The temperature of the mixture is $T=x_1T_1+x_2T_2$, where 
$x_i=n_i/(n_1+n_2)$ is the mole fraction of species $i$. The balance 
equation of the granular temperature for species $i$ can be obtained by 
multiplying the Boltzmann equations (\ref{1}) by $m_iV^2$ and integrating 
over ${\bf V}$. The result is   
\begin{equation}
\label{4}
aP_{i,xy}+\frac{d}{2}\zeta_i p_i=0,
\end{equation}
where $p_i=n_iT_i$, 
\begin{equation}
\label{5}
{\sf P}_i=m_i\int d{\bf V} {\bf V}{\bf V}f_i({\bf V}),
\end{equation}
is the partial pressure tensor of the species $i$ and 
$\zeta_i=\sum_j\zeta_{ij}$ is the cooling rate for the partial temperature 
$T_i$, with
\begin{equation}
\label{6}
\zeta_{ij}=-\frac{1}{dn_iT_i}\int 
d{\bf V}m_{i}V^{2}J_{ij}[{\bf V}|f_{i},f_{j}]\;.
\end{equation} 
According to Eq.\ (\ref{4}), in the steady state the temperature ratio 
$\gamma\equiv T_1/T_2$ is given by the relation
\begin{equation}
\label{7}
\gamma=\frac{x_2\zeta_2P_{1,xy}}{x_1\zeta_1P_{2,xy}}.
\end{equation}
Thus, to get $\gamma$ one needs to determine the cooling rates $\zeta_i$ and 
the $xy$ element of the partial pressure tensors ${\sf P}_i$. An equation 
for the elements of ${\sf P}_i$ follows immediately from the definition 
(\ref{5}) and the Boltzmann equation (\ref{1}):
\begin{equation}
\label{8}
a_{km}P_{i,m\ell}+a_{\ell m}{P}_{i,mk}=\sum_j {A}_{ij,k\ell}, 
\end{equation}
where we have introduced the collisional moments ${\sf A}_{ij}$ as 
\begin{equation}
\label{9}
{\sf A}_{ij}=m_i\int d{\bf V} {\bf V}{\bf V} J_{ij}[{\bf V}|f_i,f_j].
\end{equation}
The determination of ${\sf P}_i$ is a closed problem provided the moments 
${\sf A}_{ij}$ are explicitly known. This requires the knowledge of the 
velocity distribution functions $f_i$, which is quite an intricate problem, 
even in the elastic case. A useful way to estimate $\zeta_{i}$ and 
${\sf A}_{ij}$ is to expand $f_i$ in Sonine polynomials. This approach is 
similar to the usual moment method for solving kinetic equations in the 
elastic case where the leading order truncation is known to be a good 
approximation. In the case of 
shear flow, we take the leading Sonine approximation, $f_i({\bf V})\to 
f_{i,M}({\bf V})\left[1+{\sf C}_i:{\sf D}_i({\bf V})/2T_i\right]$, where 
${\sf C}_i=({\sf P}_i/p_i)-\openone$ and ${\bf D}_i({\bf V})=m_i
\left[{\bf V}{\bf V}-(V^2/d)\openone\right]$. Here, $\openone$ is the 
$d\times d$ unit tensor and $f_{i,M}$ is a Maxwellian distribution at the 
temperature of the species $i$, i.e.,
\begin{equation}
\label{11} 
f_{i,M}({\bf V})=n_i \left(\frac{m_i}{2\pi 
T_i}\right)^{d/2}\exp\left(-\frac{m_iV^2}{2T_i}\right). 
\end{equation} 
With this approximation, the integrals appearing in the 
expressions of the cooling rates $\zeta_{ij}$ and the collisional moments 
${\sf A}_{ij}$ can be explicitly evaluated. Retaining only linear terms 
in ${\sf C}_i$ and after a lengthy calculation, one gets 
\begin{equation}
\label{12} 
\zeta_{ij}=\frac{2\pi^{(d-1)/2}}{d\Gamma(d/2)}n_j\mu_{ji}\sigma_{ij}^{d-1}
v_0 \left(\frac{\theta_i+\theta_j}{\theta_i\theta_j}\right)^{1/2}
(1+\alpha_{ij})\left[2-\mu_{ji}(1+\alpha_{ij})
\frac{\theta_i+\theta_j}{\theta_j}\right],
\end{equation}
\begin{eqnarray}
\label{13}
{\sf A}_{ij}&=&-\frac{\pi^{(d-1)/2}}{d\Gamma(d/2)}m_i
n_in_j\mu_{ji}\sigma_{ij}^{d-1}v_0^3 
\left(\frac{\theta_i+\theta_j}{\theta_i\theta_j}\right)^{3/2}
(1+\alpha_{ij})\left\{\left[\lambda_{ij}-\frac{d}{d+3}\mu_{ji}(1+\alpha_{ij})
\right]\openone \right. \nonumber\\
& & \left. 
+2\frac{\theta_i\theta_j}{(\theta_i+\theta_j)^2}\left[\left(1+
\frac{d+3}{2(d+2)}\frac{\theta_i+\theta_j}{\theta_i}\lambda_{ij}\right){\sf 
C}_i-\left(1-\frac{d+3}{2(d+2)}
\frac{\theta_i+\theta_j}{\theta_j}\lambda_{ij}\right){\sf 
C}_j\right]\right\}.
\end{eqnarray}
In these expressions, $v_0=\sqrt{2T(m_1+m_2)/m_1m_2}$ is a thermal velocity 
defined in terms of the temperature of the mixture $T$, and 
\begin{equation}
\label{14}
\theta_1=\frac{1+x_1(\gamma-1)}{\mu_{21}\gamma}, \quad  
\theta_2=\frac{1+x_1(\gamma-1)}{\mu_{12}},
\end{equation}
\begin{equation}
\label{15}
\lambda_{ij}=2\frac{\mu_{ij}\theta_j-\mu_{ji}\theta_i}{\theta_i+\theta_j}+
\frac{\mu_{ji}}{d+3}(2d+3-3\alpha_{ij}).
\end{equation} 
Equations (\ref{12}) and (\ref{13}) extend previous expressions\cite{MG02} 
obtained in the three dimensional case. The approximation (\ref{13}) allows 
one to solve the set of equations (\ref{8}) and express ${\sf P}_i$ as a 
function of $\gamma$, while the approximation (\ref{12}) gives the cooling 
rates $\zeta_i$. When all these expressions are used in Eq.\ (\ref{7}), one 
gets a {\em closed} equation for the temperature ratio $\gamma$, that can be 
solved numerically.

A full presentation of the results is not possible because of the complexity 
of the parameter space: $\alpha_{11}$, $\alpha_{22}$, $\alpha_{12}$, 
$m_1/m_2$, $x_1$, and $\sigma_1/\sigma_2$. As in Ref.\ \cite{CH02}, we 
assume for the sake of illustration that the spheres or disks are made of 
the same material ($\alpha_{ij}=\alpha$) and have the same mass density, 
i.e., $m_1/m_2=(\sigma_1/\sigma_2)^{d}$. This reduces the parameter space 
to $\alpha$, $x_1$, and $\sigma_1/\sigma_2$. As expected, our results show 
that in general the kinetic temperatures of the mixture are different 
($\gamma\neq 1$). There are only two trivial exceptions: the elastic case 
($\alpha=1$) and the case of mechanically equivalent particles ($m_1=m_2$, 
$\sigma_1=\sigma_2$). Beyond these cases, the dependence of $\gamma$ on the 
parameters of the problem is quite intricate. As an illustration, we plot 
the temperature ratio $T_1/T_2$ versus the diameter ratio 
$\sigma_1/\sigma_2$ for $x_1=\case{1}{3}$ (Fig.\ \ref{fig1}) and 
$x_1=\case{1}{2}$ (Fig.\ \ref{fig2}) for three different values of $\alpha$: 
$\alpha=0.95$, $\alpha=0.9$, and $\alpha=0.8$. We have considered the two 
dimensional case ($d=2$). Also, for comparison we show the kinetic-theory 
predictions of Jenkins and Mancini\cite{JM87}. It is apparent that 
an excellent agreement between Monte Carlo simulations (symbols) and our 
theory is found over the entire range of values of size and mass ratios 
considered. Although the solid fractions considered by Clelland and 
Hrenya\cite{CH02} prevent us from making a quantitative comparison between 
our theory and their simulations, we observe that the behavior of $\gamma$ 
in dilute systems is qualitatively similar to the one found in Ref.\ 
\cite{CH02}. Thus, for instance, at a given value of $\alpha$ the granular 
energy of the larger particle (say for instance, species 1) increases 
relative to that of the smaller particle as the size ratio 
$\sigma_1/\sigma_2$ increases. Both Monte Carlo simulation and theory show 
that the temperature ratio presents a strong dependence on the restitution 
coefficient. With respect to the influence of composition, comparison of 
Figs.\ \ref{fig1} and \ref{fig2} indicates that $\gamma$ exhibits a very 
weak dependence on the mole fraction $x_1$. This behavior has been also 
found in recent experiments \cite{FM02} carried out on binary vibrated 
granular gases. We also see that all the above trends are qualitatively 
reproduced by the theory of Jenkins and Mancini \cite{JM87} (which is 
restricted to nearly elastic disks), although these trends are however 
strongly exaggerated. Thus, for instance, at a size ratio 
$\sigma_1/\sigma_2=3$, in the case $x_1=\case{1}{2}$ and $\alpha=0.8$, the 
discrepancy between our theory and the simulation is less than $1\%$ while 
it is around of $115\%$ in the theory of Jenkins and Mancini \cite{JM87}.

An interesting point is to assess the influence of the temperature 
differences on the rheological properties of the mixture. Although the 
comparison carried out in Ref.\ \cite{CH02} between previous theories (based 
on a single temperature) and simulation shows a qualitative good agreement 
at the level of the shear stresses, one expects that the new contributions 
coming from the energy difference leads to an improvement over previous 
theoretical predictions. In Fig.\ \ref{fig3}, we plot the dimensionless 
stresses $-P_{xy}^*$ and $P_{yy}^*$ versus the restitution coefficient in 
the case $x_1=\case{1}{2}$ and $m_1/m_2=(\sigma_1/\sigma_2)^{1/2}=10$. Here, 
${\sf P}^*={\sf P}/nT=({\sf P}_1+{\sf P}_2)/nT$. Also shown in this Fig.\ 
\ref{fig3} is the result which would be obtained if the differences in the 
partial temperatures were neglected [i.e.,  $\theta_i=\mu_{ji}^{-1}$ in 
Eqs.\ (\ref{12}) and (\ref{13})]. In general, inclusion of the 
two-temperature effects represents a significant improvement of the theory, 
especially in the case of the shear stress $P_{xy}^*$, which is the most 
relevant rheological property in a sheared flow problem. This justifies the 
use of a two-temperature description to capture the dependence of stresses 
on dissipation.

In summary, we have obtained an approximate evaluation of the 
temperature ratio of a sheared granular mixture. The accuracy of this 
calculation has been also confirmed by Monte Carlo simulation of the 
Boltzmann equation. As was also found in recent MD simulations \cite{CH02}, 
our results show that the temperature ratio strongly depends on dissipation 
and the mechanical parameters of the mixture (especially on the ratios of 
mass and size). In addition, the effect of temperature differences on 
rheology is important (especially in the case of the shear stress) and leads 
to an improvement of the theoretical results with respect to the predictions 
made from a single-temperature theory.

V. G. acknowledges partial support from the Ministerio de Ciencia y 
Tecnolog\'{\i}a (Spain) through Grant No. BFM2001-0718.

\begin{figure}
\caption{Plot of the temperature ratio $T_1/T_2$ as a function of the size 
ratio $\sigma_1/\sigma_2=(m_1/m_2)^{1/2}$ for a two-dimensional system in 
the case $x_1=1/3$ and three different values of the restitution coefficient 
$\alpha$: (a) $\alpha=0.95$, (b) $\alpha=0.9$, and (c) $\alpha=0.8$. The 
solid lines are the theoretical predictions while the symbols refer to the 
Monte Carlo simulation results. The dashed line corresponds to the 
prediction given by the theory of Jenkins and Mancini \protect{\cite{JM87}} 
in the case $\alpha=0.8$.
\label{fig1}}
\end{figure}
\begin{figure}
\caption{The same as in Fig.\ \protect{\ref{fig1}} but for $x_1=1/2$.
\label{fig2}}
\end{figure}
\begin{figure}
\caption{Plot of the reduced elements of the pressure tensor 
$P_{yy}^*=P_{yy}/nT$ and $P_{xy}^*=P_{xy}/nT$ as a function of the 
restitution coefficient $\alpha$ for a two-dimensional system in 
the case $m_1/m_2=10$ and $x_1=1/2$. The solid lines are the theoretical 
predictions while the symbols refer to the Monte Carlo simulation results. 
The dashed lines correspond to the theoretical results by assuming the 
equality of the partial temperatures $T_1/T_2=1$.
\label{fig3}}
\end{figure}

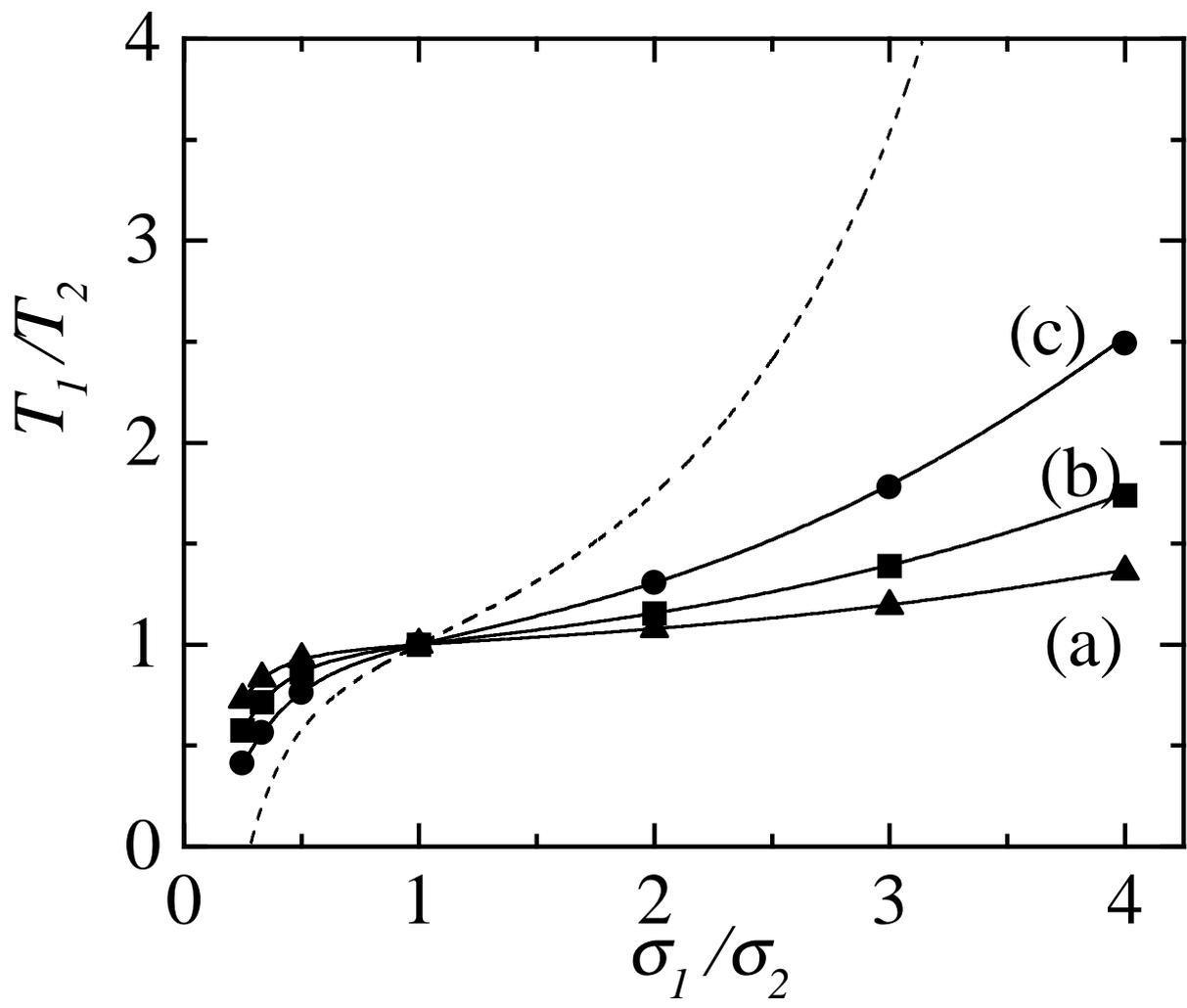
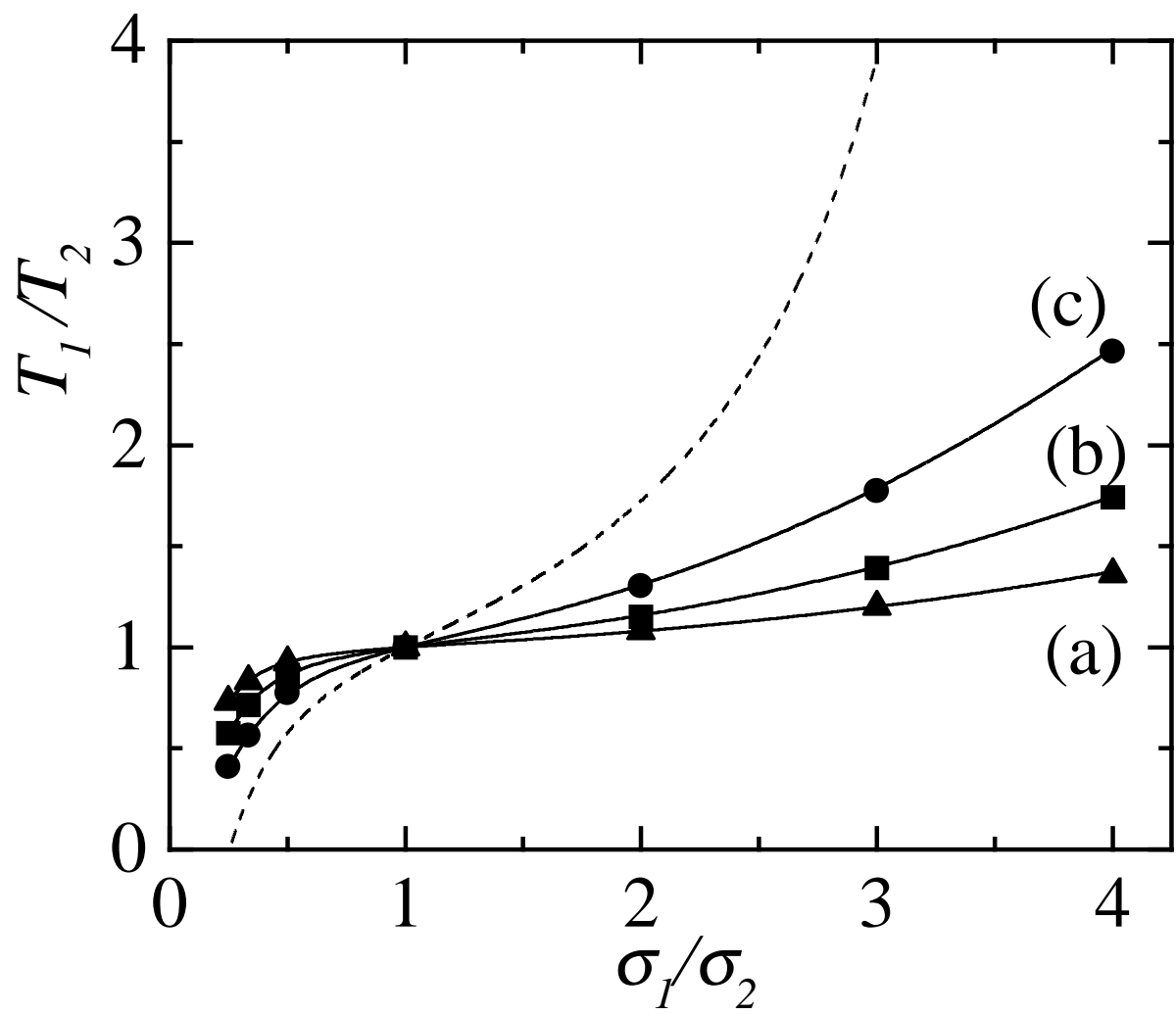
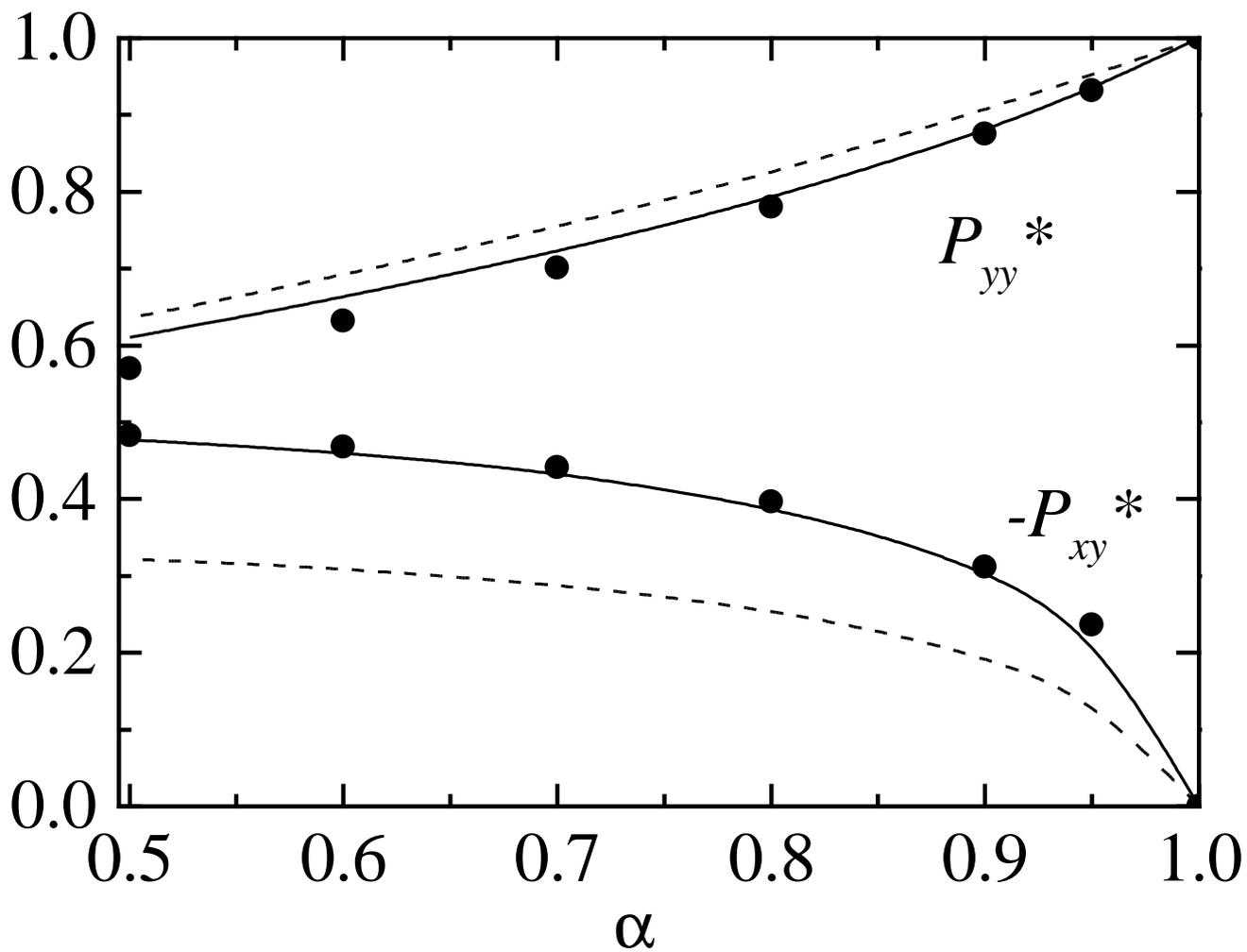
\end{document}